# Planar Aperiodic Arrays as Metasurfaces for Optical Near-Field Patterning

*Mario Miscuglio[1,2], Nicholas J. Borys[3], Davide Spirito[1], Beatriz Martín-García[1], Remo Proietti Zaccaria[1], Alexander Weber-Bargioni[3], P. James Schuck[3], Roman Krahne[1]\**

[1]Nanochemistry Department, Istituto Italiano di Tecnologia, Via Morego 30, 16163 Genova, Italy

[2]Dipartimento di Chimica e Chimica Industriale, Università degli Studi di Genova, Via Dodecaneso, 31, 16146, Genova, Italy.

[3]Molecular Foundry, Lawrence Berkeley National Lab, 1 Cyclotron Road, Berkeley, California 94720, USA

ABSTRACT Plasmonic metasurfaces have spawned the field of flat optics using nanostructured planar metallic or dielectric surfaces that can replace bulky optical elements and enhance the capabilities of traditional far-field optics. Furthermore, the potential of flat optics can go far beyond far-field modulation, and can be exploited for functionality in the near-field itself. Here, we design metasurfaces based on aperiodic arrays of plasmonic Au nanostructures for tailoring the optical near-field in the visible and near-infrared spectral range. The basic element of the arrays is a rhomboid that is modulated in size, orientation and position to achieve the desired functionality of the micron-size metasurface structure. Using two-photon-photoluminescence as a tool to probe the near-field profiles in the plane of the metasurfaces, we demonstrate the molding of light into different near-field intensity patterns and active pattern control *via* the far-field illumination. Finite element method simulations reveal that the near-field modulation occurs *via* a combination of the plasmonic resonances of the rhomboids and field enhancement in the nanoscale gaps in between the elements. This approach enables optical elements that can switch the near-field distribution across the metasurface *via* wavelength and polarization of the incident far-field light, and provides pathways for light matter interaction in integrated devices.

The phase, amplitude, directionality, dispersion and polarization of light can be modulated with subwavelength metal antennas,[1, 2] which gives rise to the field of plasmonic metasurfaces. The collective effects of an ensemble of nanoscatterers, for example of V-shaped nanoantennas as introduced by Yu *et al.*,[3] can mold the wavefront of the incident light, and shape the properties of the reflected and refracted beams.[4, 5] This concept has proven to be a powerful approach to replace conventional bulk optical elements such as lenses and polarizers with ultrathin planar surfaces that are coated with periodic or aperiodic arrays of metallic or dielectric nanostructures.[6] The light modulation occurs in the near-field at the interface with the nanostructured surface, and most efforts—as illustrated in Figure 1A (i)—have been successfully directed to establish far-field optical elements with functionalities such as achromatic focusing, polarization manipulation, and near-perfect absorbtion/reflection,[7-17] which is a rapidly growing field. Near-field optical characterization of these devices aims to capture fundamental properties of such planar metasurfaces.[18,19] Furthermore, planar metasurfaces can be used to couple far-field radiation to plasmons,[20] for example by designing arrays of subwavelength apertures in a metallic film that launch propagating plasmon polaritons at a metal-dielectric interface.[21] And metasurfaces can be used to modify the radiation from near-field sources and for waveguiding of plasmons in thin film structures, as discussed in a recent review.[22] The concept of nanostructured flat optics to engineer optical elements that act in the plane or vicinity of the metasurface is illustrated in Figure 1A(ii). In this respect, the ability of planar metasurfaces to modulate the near-field properties provides a plethora of opportunities that is not limited by the generation propagating plasmons.[23]

By using the coupling of localized nanoscale metallic elements in specifically designed structures, the light intensity in the near-field can be molded into patterns for a targeted functionality. One simple and extremely successful manifestation is a bowtie dimer of two gold nanoantennas that generates a strongly confined hot spot.[24] One could imagine to extend this concept to larger arrays that can achieve a variety of functionalities on a larger scale. For example, mimicking how spherical apexes are used to couple effectively the far-field to the near-field used in tip-enhanced Raman spectroscopy (TERS)[25-27] or scattering near-field optical microscopy (SNOM),[28-31] flat metasurfaces may enable efficient interfacing of light to nanoscale optoelectronic elements. The modulation of amplitude and phase by arrays of plasmonic nanoresonators could be directly exploited to shape the near-field, and thereby designing concentrators, switches or optical traps that act in the vicinity of the metasurface itself. In this respect, deterministic aperiodic patterns of plasmonic nanostructures have been proven to be a viable approach for optical near-field modulation and the creation of complex scattering resonances.[32-34]

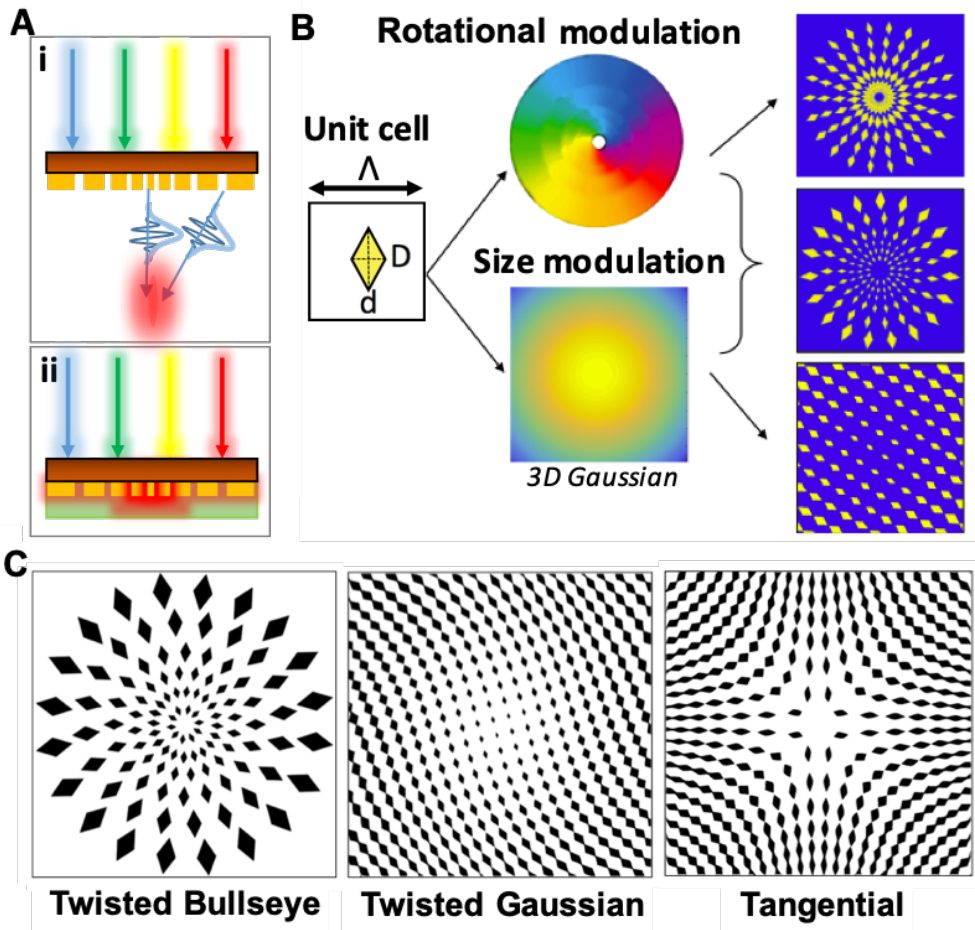

**Figure 1. Concept of functional near-field optics and metasurface pattern design. (A)** Illustration of the difference of flat plasmonic metasurfaces for far-field optics (i) and near-field functionality in thin film technology (ii). **(B)** Scheme of the design method used in this work to achieve aperiodic arrays with specific functionality. The size and orientation angle of the unit element (here a rhombus) are varied by a mathematical algorithm to obtain a modulation of the pattern, resulting in a specific distribution of the optical near-field. **(C)** Schemes of the three functional patterns obtained by different design approaches: the Twisted Bullseye (TBE), the Twisted Gaussian, and the Tangential pattern.

In this work, we fabricate aperiodic arrays of nanoresonators to achieve a set of diverse functionalities that include light concentration at the center of the pattern, generation of optical vortices, and switching of the near-field intensity pattern controlled by the wavelength or linear polarization angle of the far-field illumination. With a specific functionality in mind, we design the patterns of plasmonic nanoscatterers using a mathematical algorithm based on Fast Fourier Transform (FFT) that achieves the targeted modulation, for example, by tuning size and orientation of the elements of the array, as depicted in Figure 1B. To illustrate the broad versatility range of this approach, we chose three different routes for the design of the patterns: one based on modulation of an array in radial coordinates (Figure 1C(i)); a second based on modulating the Cartesian coordinates (Figure 1C(ii)); and a third that achieves modulation by the projection of a mathematical function (Figure 1C (iii)). The metasurface structures were fabricated by electron beam lithography and their optical properties were subsequently investigated. For the optical characterization, it is critical to characterize the response of the structures using spatially homogenous, widefield illumination since the light modulation is a collective effect of the rhomboidal plasmonic resonators that constitute the metasurface patterns. For far-field investigation, we used polarized optical microscopy and defocused dark field microscopy. To gain insight into the near-field distribution of the light while maintaining widefield illumination, we used widefield two-photon photoluminescence (TPPL) microscopy. By imaging upconverted, nonlinear emission from the gold structures, the TPPL experiments probe the optical near-field and reveal its redistribution induced by the patterns. Finite element simulations of the electrical near-field show the plasmonic resonances of the elements, corroborate our experimental characterization, and provide insight into the coupling among the individual elements in the aperiodic arrays.

## Results/Discussion

The basic element of the metasurface patterns is a rhombus made from Au with an aspect ratio of 2 and lateral dimensions that range from a few tens to some hundreds of nm, and a thickness of around 30 nm. We chose the rhombus since it can sustain pronounced dipolar and higher plasmonic resonances, manifesting fundamental modes at different frequencies in the orthogonal directions of the two major axes. Furthermore, the tip shape at the extremities favors the plasmonic coupling to neighboring elements, which is crucial for our approach. The amplitude of the electric field for the fundamental and first overtone resonant eigenmodes of the rhombus in $x$- and $y$-direction are shown in the Supporting Information (SI) in Figure S1A (throughout this paper, $x$ and $y$ directions refer to the in-plane horizontal and vertical directions in the images, respectively). The fundamental modes are dipolar, and the overtones are quadrupoles, and all of them demonstrate strong fields at the tips of the rhombus in the direction of the polarization. As typical for plasmonic nanoantennas, the resonance frequencies scale with size,[35, 36] and therefore a large spectral region in the near-infrared can be covered with the structures used in our design, ranging from 0.7 to 1.4 μm. We note that for practical applications the lower wavelength limit is determined by the resolution of the electron-beam lithography that was used to fabricate the structures, where we achieved well-defined rhomboidal shapes with few tens of nm in size.

Functional patterns of arrays with the rhomboids as basic elements can be designed with a deterministic mathematical algorithm in order to achieve the desired functionality of the metasurface. We aimed at three conceptually different aperiodic patterns to demonstrate the versatility of this approach: (i) the Twisted Bullseye (TBE) displayed in Figure 1C(i) is based on a radial coordinate system and designed to twist the polarization of a linearly polarized incident beam, and to concentrate the light intensity at the center of the pattern. (ii) the Twisted Gaussian in Figure 1C(ii) is an aperiodic array in Cartesian coordinates that generates vorticity, and focuses the light for a fixed angle of linear polarization, while for the perpendicular angle the near-field is located at the edges of the structure, which can be directly employed for use as optical tweezers. (iii) the Tangential pattern in Figure 1C(iii) is based on the mathematical function of tangential planes (see SI and Figure S2 for details), and leads to wavelength dependent rotation of the near-field intensity distribution.

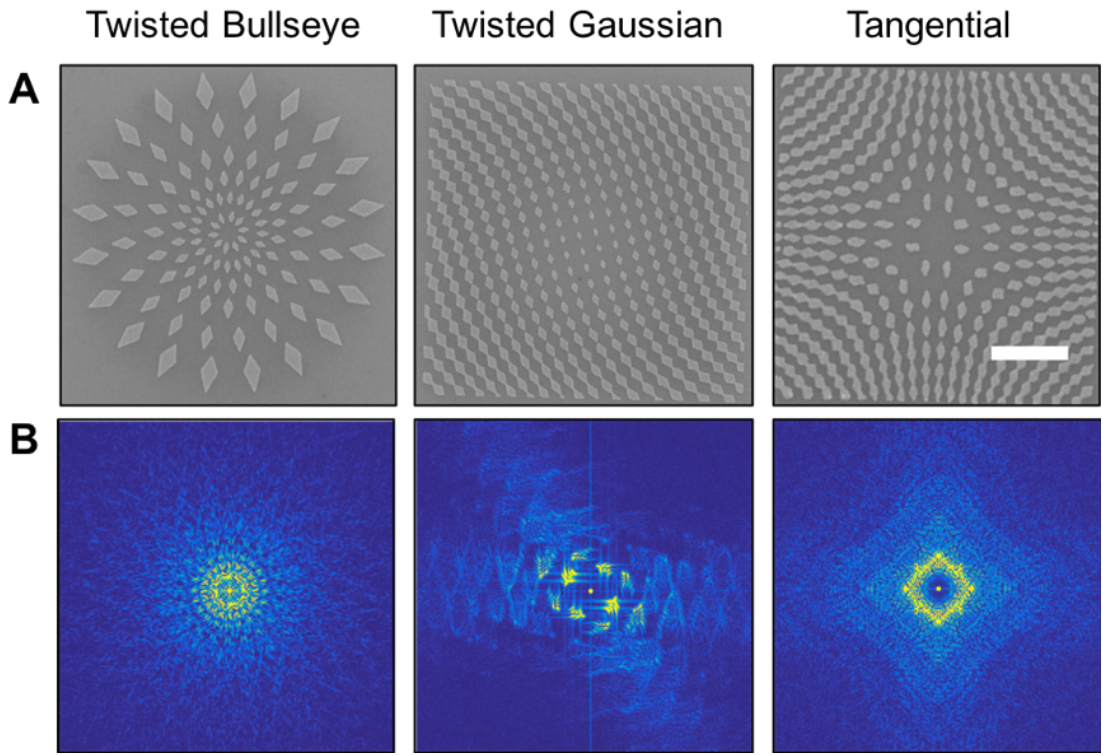

**Figure 2. Aperiodic metasurfaces investigated in this work. (A)** SEM micrographs of the experimentally fabricated Twisted Bullseye, Twisted Gaussian, and Tangential aperiodic arrays, together with their diffraction patterns calculated by FFT **(B)**. The scale bar corresponds to 500 nm and applies to all images in (A).

Figure 1B illustrates how the metasurface patterns are obtained: we start from a regular concentric or square array of rhomboids and apply a combination of modulation in element size, orientation, or shift in position. These spatial variations of the arrays can be designed such that the end result is an aperiodic structure that modulates the polarization and/or the intensity distribution of the light induced optical near-field. For example, to obtain the TBE structure, the combined modulation of orientation and size (central design in Figure 1B) leads to rhomboidal elements that are organized in concentric rings, with increasing element size from the center outwards, and a twist in orientation between adjacent rings of elements by an angle of 30 °. The detailed underlying mathematical approach is reported in the Supporting Information. SEM images of the TBE, Twisted Gaussian, and Tangential metasurface patterns designed by this algorithm are shown in Figure 2A, and their FFT calculated diffraction patterns are displayed in Figure 2B. For the TBE and the Twisted Gaussian, the diffraction patterns displayed in the lower panels do not show mirror symmetry along the *x*- and *y*-axes, and therefore are characterized by a notable vorticity. Furthermore, the Bragg points in the diffraction patterns are surrounded by satellites, which is a signature of aperiodic, phase modulated crystals.[37] Different from the TBE and the Twisted Gaussian, the Tangential pattern in Figure 1C(iii) maintains the mirror symmetry, and does not introduce vorticity.

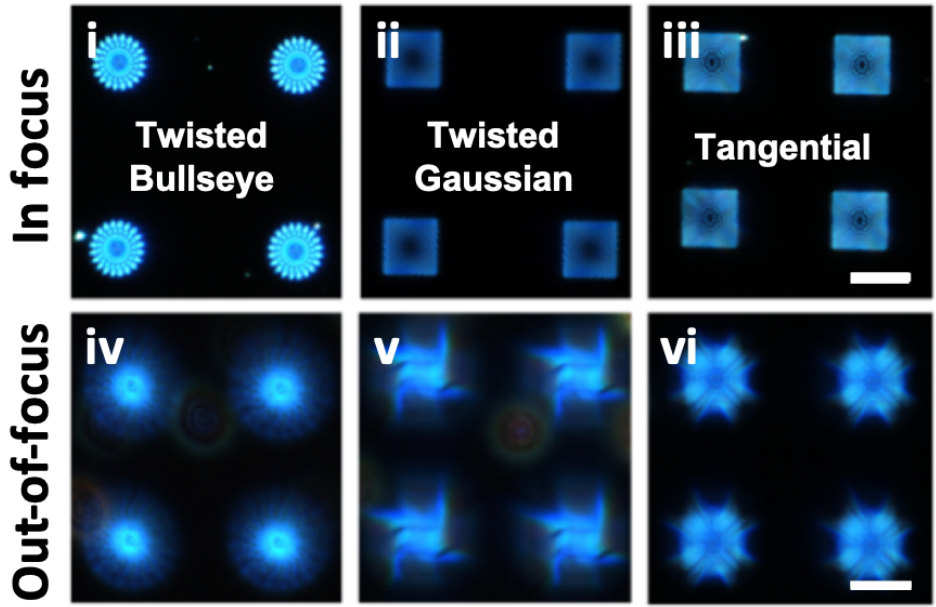

**Figure 3. Far-field response of the aperiodic metasurface patterns.** Optical dark field images in- and out-of focus (with focus of ca. 3 µm away from the surface) recorded from arrays of the three functional patterns. The far-field modulation of the plasmonic patterns is manifested in the out-of-focus images by a vortex for the TBE and Twisted Gaussian patterns, and by the 4 leaf-clover like light distribution for the Tangential pattern. Scale bars are 5 µm and apply to all images.

Figure 3 shows microscopic dark field images that are in- and out-of-focus of 2x2 arrays of the structures. In focus, the design of the pattern can be distinguished, and the light intensity reflects the scattering from the elements. The out-of-focus images, which give information about the spatial distribution of the light intensity in the vicinity of the structures,[38-40] manifest a completely different behavior. For the TBE, the maximum intensity is in a donut close to the center of the rings, and clearly a vortex-like distribution spreading outwards can be recognized. Vorticity is also strongly present in the out-of-focus image of the Twisted Gaussian pattern, where the light intensity resembles a four-leaved pinwheel toy. The out-of-focus image of the Tangential pattern in Figure 3B shows that light is diverged towards the four corners of the pattern, making the hyperbolic intensity lines clearly identifiable. These optical microscopy images already indicate the near-field modulation induced by the metasurfaces that penetrates into the far-field, altering light polarization and field distribution. Polarized microscopy experiments in bright and dark field recorded from the TBE structures are shown in Figure S3 and confirm the projected polarization rotation.

Augmenting the far-field characterization discussed above, we performed TPPL microscopy experiments to image and directly measure the spatial distribution of the optical near-field modulation generated by our structures. The TPPL signal is proportional to the square of the light intensity (see Figure S4), and therefore extremely sensitive to the local near-field enhancement.[41-46] The TPPL emission itself arises from the nonlinear excitation and subsequent inelastic emission of PL from the Au elements in the arrays. Therefore, the TPPL of the metasurface structures is a convolution of the 4$^{th}$ power of the local electric field distribution, the presence of Au material, and the diffraction-limited instrument response function of the wide-field imaging system.[41, 47, 48] The Experimental section and the SI with Figure S5 provide details on the TPPL microscopy setup. Figure 4A shows the TPPL intensity map for a TBE metasurface with a 1.25 μm diameter for different excitation wavelengths of linear polarized light in the *y* direction. For excitation at 700 nm, two clearly defined lobes on the horizontal axis are present, whereas for 800 nm, a strongly localized emission hotspot is observed at the center of the pattern with a diffraction limited full-width at half maximum (FWHM) of around 300 nm with nearly fully diminished side lobes. Finite element method simulations (COMSOL Multiphysics) that compute the 4$^{th}$ power of the norm of the electrical near-field convoluted with a 2D Gaussian function (to account for the diffraction-limited imaging) reproduce this behavior well, as demonstrated in Figure 4B. The computed distribution of the norm of the electrical nearfield in Figure 4C gives deeper insight into the underlying mechanisms of the light concentration that is induced by the TBE array. At 700 nm excitation, the rhomboidal elements manifest hot spots at their corners that correspond to their short and long axis dipolar and quadrupolar resonances (see also Figure S1). In particular, horizontally oriented rhomboids show strong hot spots at their flat corners that correspond to the short axis plasmonic resonances. These strong resonances lead to the horizontal lobes observed by TPPL at 700 nm excitation. For 800 nm excitation the relative intensity of these short axis resonances is reduced, and we observe a strong nearfield amplitude in the tip-to-tip gaps of the rhomboids in adjacent rings. This behavior resembles that of bowtie antennas, and the distribution of this nearfield increases in the gaps becomes more dominant and circular towards the center, where small gaps and small structures are present. This trend is in good agreement with the reports on bowtie antennas.[24, 47] The TBE array can be seen as a self-similar circular structure, and indeed the near-field distribution in Figure 4C at 800 nm resembles that of self-similar antennas if one follows the rhomboid chains from outside to the center of the pattern.[49, 50] This effect is nicely reflected by the strong near-field intensity at the center of the pattern where no gold structures are present, but which represents the "focal point" of the self-similar lens represented by the TBE. The correspondence of calculated cross section with the integrated TPPL intensity confirms that the TPPL intensity can be used as a measure for the optical near-field, as shown in Figure S7.

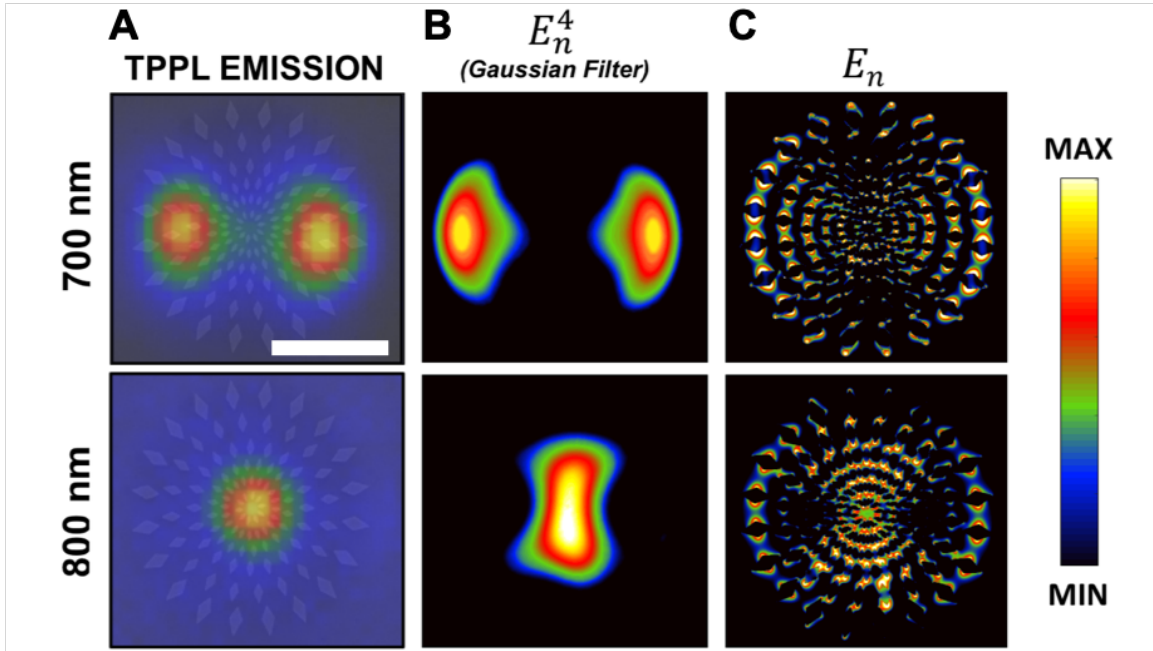

**Figure 4. Near-field modulation of the TBE pattern. (A)** Light intensity maps recorded by widefield two-photon-photoluminescence (TPPL) imaging of the same TBE structure with 1.25 µm diameter under excitation at 700 and 800 nm. The TPPL maps are overlayed with the SEM image of the pattern. **(B)** The 4$^{th}$ power of the calculated electric near-field, $E_n = \sqrt{E_X^2 + E_Y^2 + E_Z^2}$, of the TBE pattern convoluted with 2D Gaussian function with 300 nm full width at half maximum (FWHM). **(C)** Calculated electric near-field, $E_n$, of the TBE pattern evidencing the coupling of the elements that focuses the electric field towards the center of the structure. Scale bar corresponds to 500 nm and applies to all panels.

Structures with different size were obtained by linearly scaling the whole pattern, and Figure S9 reports the element size range *versus* the pattern diameter. TPPL images recorded at different excitation wavelengths and for structures with different sizes (see Figure S8) show that the modulation effect of the TBE structures scales with size and wavelength. Larger structures are resonant with longer wavelength, and thus the concentration of the emission to a circular spot at the center of the pattern is shifted to longer wavelengths as compared to smaller structures.

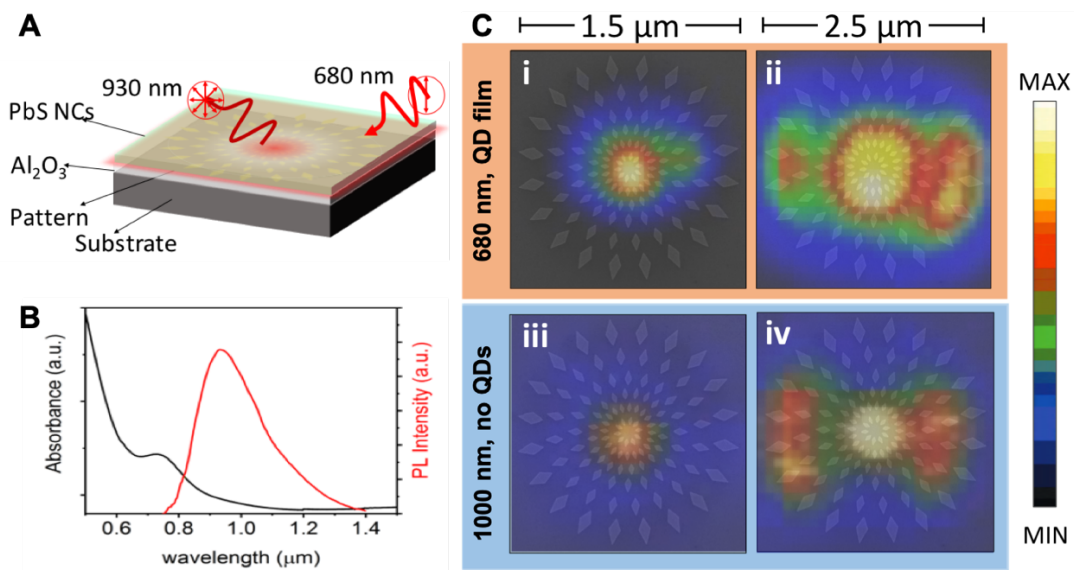

**Figure 5. Modulation of the emission of a PbS nanocrystal film with the TBE pattern. (A)** Illustration of the device structure: A thin layer of $Al_2O_3$ and a film of PbS nanocrystals were sequentially deposited on the TBE pattern made from Au. The linearly polarized excitation at 680 nm wavelength and the NC film emission at 930 nm are sketched by the red arrows. **(B)** Absorption and photoluminescence spectra of the PbS nanocrystal film. **(C)** TPPL intensity maps of TBE patterns with different size under linearly polarized excitation at 680 nm for devices with a nanocrystal film, and at 1000 nm excitation for devices without a nanocrystal film coating (see also the profiles for 950 nm excitation in the Figure S8A). For a given structure size, the TPPL intensity profile is very similar for the two cases.

To test if the light modulation effect can be transferred to the emission of fluorophores deposited on top of the plasmonic metasurfaces, we coated the TBE patterns with a thin film of PbS nanocrystals (NCs), as illustrated in Figure 5A. Here the conceptual idea is that the incident light at short wavelength is strongly absorbed by the NC film, leading to emission from the NC film at a longer wavelength that corresponds to the excitonic band gap of the NC material. This mechanism corresponds to down-conversion of the incident light. The interesting point is if the emission from a film directly deposited on the metasurface pattern has the same effect as illumination with a homogeneous beam from a light source that is far from the substrate surface. For the sample fabrication, first, a thin film of alumina (10 nm) was deposited on top of the sample surface to avoid metal-induced quenching of the nanocrystal emission. The alumina deposition was then followed by spin-coating of the PbS NCs with a diameter of 2.2 nm (dispersed in toluene, see Figure S10), resulting in a film with a thickness of ~20 nm. The PbS nanocrystal films have an absorption onset at ~800 nm, and an emission peak centered at 930 nm (see Figure 5B). Figure 5C compares the TPPL intensity maps of structures with different size that were coated with the PbS nanocrystal film excited at 680 nm, with those of the bare metasurface patterns excited at 1000 nm. If the metasurface responds in both cases to the incident light, the modulation should be very different due to the different excitation wavelength (see Figure 4). However, if the metasurface interacts with the down-converted PbS nanocrystal emission at 930 nm, then similar near-field distributions in both cases for a given structure size would be expected. Figure 5C unambiguously shows the latter behavior and confirm that nanocrystal film coated TBE structures interact with the down-converted emission of the nanocrystal film at 930 nm. Therefore, the modulation effect of the near-field can be transferred to the emission of fluorophores deposited on top of the metasurface, which enables interesting opportunities for the use of fluorophore coated metasurfaces in multiplexing. Here different combinations of patterns of different sizes and functionalities could be combined with fluorophores that emit at different wavelengths, but the whole device could be illuminated off-resonance by a single light source at shorter wavelength.

Now we discuss the Twisted Gaussian pattern based on Cartesian coordinates that due to the aspect ratio of the rhomboid elements has a fundamentally different symmetry with respect to horizontal or vertical direction, and therefore its near-field modulation can be expected to depend on the angle of incidence of linearly polarized light. Figure 6 shows the TPPL intensity maps for the Twisted Gaussian pattern at different excitation wavelengths, recorded with linearly polarized light in the *x*-direction. The intensity profile shows a donut shape that is well focused to the center for the shorter wavelengths, and which gets larger for longer wavelengths. As expected, the TPPL distribution for the Twisted Gaussian pattern depends strongly on the angle of the incident linear polarization, as shown in Figure 6B. For linear polarization in the *y*-direction we obtain a strongly focused circular spot at the center, whereas for polarization in the *x*-direction, the intensity distribution resembles a large donut shape, with higher intensity on the left and right sides. The Twisted Gaussian metasurface therefore can be used to switch the near-field intensity distribution *via* the polarization angle of the incident light (Figure 6B).

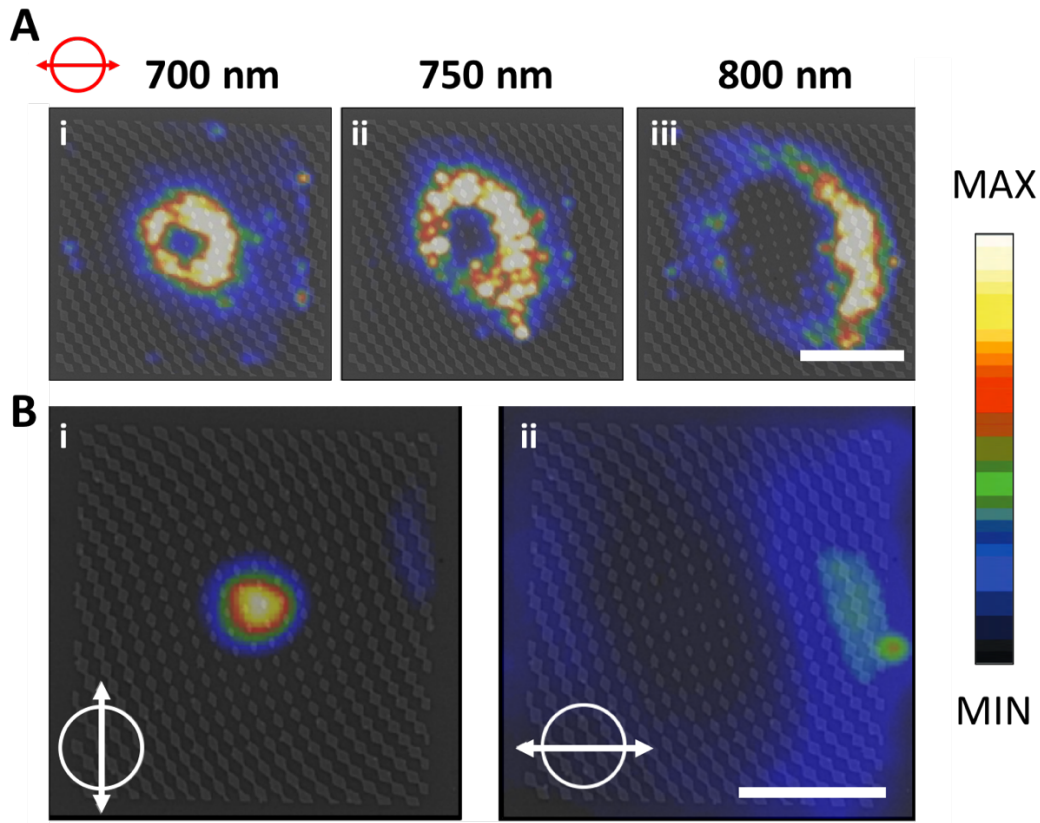

**Figure 6. Twisted Gaussian pattern: Wavelength and polarization dependent modulation of the optical near-field. (A)** TPPL signal of the Twisted Gaussian patterns obtained under horizontally polarized excitation at different wavelengths. The diameter of the donut shaped near-field intensity increases with increasing excitation wavelength. **(B)** Rotation of the linear polarization of the incident light: Horizontal polarization leads to a donut-shaped field distribution, while vertical polarization focuses the near-field into a tight central spot. Scale bars are 2 μm.

Patterns generated by mathematical functions yield structures with very specific symmetry conditions and element densities. To demonstrate the versatility of this route, we designed the Tangential pattern that has horizontal, vertical, and diagonal symmetry axes, and an increased element density in the corners (see Figure 1C(iii)). The TPPL intensity maps measured from this metasurface are depicted in Figure 7A. For short wavelength, we observe a distribution with the intensity focused into triangular zones oriented in vertical direction (top and bottom). Increasing the wavelength leads to confinement of the intensity in the four corners, and for long wavelengths the intensity is focused into triangular zones in horizontal direction (left and right). Therefore, the Tangential pattern can be employed to control the near-field profiles *via* the incident wavelength, leading to a rotation of 90° of the intensity distribution, creating a plasmonic color sorter.[51]

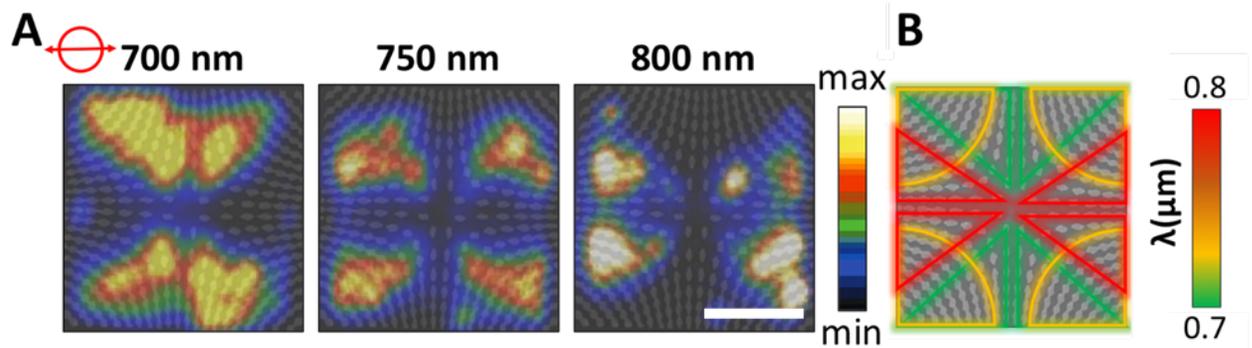

**Figure 7. Tangential pattern: Wavelength induced rotation of the near-field distribution. (A)** TPPL maps of a Tangential pattern under horizontally polarized (*x*-direction) excitation at different wavelengths. For shorter wavelengths (700 nm), a near-field distribution with two lobes that are oriented perpendicular to the incident polarization direction are obtained. At 750 nm, a distribution with light concentration in the four corners of the pattern is observed, and for 800 nm, lobes in horizontal direction appear. Scale bar is 2 μm. **(B)** Scheme illustrating the wavelength dependence of the optical near-field profiles that depicts the wavelength-dependent regions of the metasurface structure where optical energy is concentrated.

Conclusions

We introduced aperiodic metasurfaces designed with the aim to modulate the near-field distribution in the plane of the metasurface itself, and demonstrated the broad range of functionality that could be achieved with three conceptually different patterns of nanoscatterers. The functionalities include the centric concentration of light intensity, rotation of polarization, generation of optical vortices, and switching of orthogonal near-field distributions *via* illumination wavelength or polarization. TPPL experiments that probe the optical near-field of the metasurfaces reveal the collective effects of the aperiodic arrays, and finite elements simulations provide insight in the coupling between the elements. Control of the optical near-field profiles on a micron-size area is highly appealing for a broad range of applications. The focusing of the energy to a tightly confined central spot that we achieve with the TBE and Twisted Gaussian patterns should strongly enhance the coupling of the optical signal to the spherical tip apexes in TERS and in SNOM. Furthermore, the control of the optical near-field distribution with the polarization or wavelength of the incident light is particularly interesting for application such as optical tweezers. For example, with the Twisted Gaussian pattern, particles can be confined in the donut shape by horizontally polarized light, and forced out by vertically polarized light. The Tangential pattern can be used for controlling the direction of optical channels *via* the incident wavelength, with an intermediate configuration where both horizontal and vertical channels are open. Furthermore, our metasurfaces for optical near-field modulation can be employed in integrated device structures and coupled to other functional elements that are implemented at a nanoscale distance, for example to light emitting quantum dots, which allows for multiplexing devices that can function at several wavelengths in parallel. This concept exploits the near-field for information processing, counterfeit protection, and data storage.

## Methods/Experimental

Device fabrication.

Silicon substrates with 285 nm thickness of thermally grown $SiO_2$ from Graphene Supermarket® were washed in acetone (Sigma Aldrich®, CMOS grade) and then rinsed in isopropanol (Sigma Aldrich®, CMOS grade) for 2 min. The substrates were then plasma cleaned for 5 min at 100W under 100% $O_2$ for removing eventual organics, and baked at 120°C for 2 min for removing residual solvents. A layer of 495-K polymethyl-methacrylate (PMMA A2 1:1, 1% in Anisole, MicroChem®) with a thickness of 160 nm was deposited by spin-coating and baked for 7 min on a hotplate at 180°C. The nanostructure patterns were defined by electron beam lithography (Raith 150 - Two) at an accelerating voltage of 20 kV, with an exposure current of 35 pA. The resist was developed for 30s in a cold (8°C) mixture of MIBK:IPA 1:3 (Microchem®). A 5 nm thick film of Titanium (Ti pellets, 99.995%, Kurt J. Lesker®) followed by a 25 nm thick film of gold (Au pellets, 99.999%, Kurt J. Lesker®) were deposited with a rate of 0.2 Å/s through electron beam evaporation at an operating pressure of around $2\times10^{-6}$ mbar (Kenosistec Inc). Lift-off was performed in hot acetone. On the samples that were coated with a 20 nm thick film of PbS nanocrystals, a layer of $Al_2O_3$ with thickness of 10 nm was deposited by atomic-layer deposition (Flexal Oxford Instruments) prior to the spin-coating of the PbS nanocrystal film.

Optical characterization.

Dark field optical microscopy images were obtained with Nikon Eclipse LV100 microscope equipped with a 100x objective with 0.9 NA under white light. Polarized optical microscopy images were obtained with an Olympus BX41 microscope with 100x objective with 0.9 NA with white light in transmission, using two linear polarizers for the visible spectral range in rotational mounts. All images were recorded with a CCD camera.

Two-Photon Photoluminescence (TPPL) experiments.

The TPPL experiments, sample illumination and TPPL collection were performed using a 100× 0.90 NA objective lens. A pulsed Ti:sapphire laser (140 fs; 80 MHz repetition rate) with wavelengths from 680-1080 nm (with a ~4 nm bandwidth) was used as the excitation source. In order to achieve quasi-homogeneous illumination over the metasurface structures, the laser beam was expanded to a spot size with a diameter of 10 - 20 μm by focusing the laser to the back focal plane of the objective. Light collected from the sample was passed through two short-pass filters with edges at 633 and 650 nm in order to remove the excitation light from the TPPL signal. Images of the TPPL signal were recorded with a cooled scientific complementary metal oxide semiconductor camera (Andor Neo 5.5 sCMOS). The diameter of the widefield illumination spot was evaluated by the TPPL intensity from an unprocessed area of the evaporated gold film. TPPL spectra were recorded with an imaging spectrograph (Acton SP2300) and a liquid nitrogen-cooled charge coupled device (CCD) camera (Princeton Instruments SPEC-10). For the spectral measurements, a pinhole was used in an intermediate image plane to collect light from an ~1.5 μm diameter region of the sample.

Finite Elements Simulations.

The FEM simulations were performed using COMSOL Multiphysics®. For our study, we considered the metal structure on top of $SiO_2$/Si substrates. The dielectric permittivity for Au, Ti and Si was taken from Palik.[52] The plane wave radiation originated from a port placed above the structure, which allowed for specular reflection, while another port was placed below the structure which absorbed the transmitted plane wave. An outer box corresponding to a Perfectly Matched Layer (PML) was implemented for minimizing unphysical reflections of the scattered waves. The system was at first illuminated at a normal incidence with a linearly polarized plane wave (TEM) with a varying wavelength. The boundary conditions were taken as Floquet periodic conditions. We used a sufficiently fine mesh that gave steady and mesh independent results for the electric field distribution, where the largest elements of the mesh were smaller than 1/10 of the wavelength. A direct solver was chosen for the solution method (MUMPS), which allowed cluster computing. The machine used for the simulation was an Intel Xeon (X5690) at 3.47 GHz dual processor with 192 GB RAM memory.

Synthesis of PbS nanocrystals.

The synthesis of colloidal PbS NCs was performed by slight modifications of established protocols.[53] Briefly, a mixture of PbO (0.465 g, 99.999%, Alfa Aesar®), oleic acid (2.5 mL, 90%, Sigma Aldrich®) and octadecene (ODE, 90%, Sigma Aldrich®) (10 mL), was heated and degassed under vacuum at 90°C for several hours, obtaining a clear solution. Then, the flask was cooled to 80ºC and a mixture of hexamethyldisilathiane (207 μL, synthesis grade, Sigma Aldrich®) and ODE (10 mL) was rapidly injected. After 30s of crystal growth, the reaction was quenched using an ice bath in order to obtain monodisperse small size oleic acid-capped PbS NCs (ca. 2.2 nm diameter). For the preparation of the films, the PbS NC suspension as obtained from synthesis was purified twice (in air) by adding acetone, followed by centrifugation. The NCs were finally dispersed in toluene and filtered with a 0.2 μm PTFE filter membrane (Sartorius®). Finally, the PbS NCs solution (30 mg/mL) was spin coated with 2500 rpm for 30s, which resulted in film coverage with 20 nm thickness (measured with a Veeco Dektak® Profilometer). Transmission electron microscopy (TEM) images of the PbS nanocrystals were acquired with a 100 kV JEOL JEM-1011 microscope for size determination. The absorption spectrum was recorded with a Varian Cary 5000 UV-vis-NIR spectrophotometer, and the photoluminescence spectrum was measured using an Edinburgh Instruments FLS920 spectrofluorometer with excitation at 400 nm. For the optical characterization in solution the PbS NCs were dispersed in tetrachloroethylene (anhydrous, ≥99%, Sigma Aldrich®). PL maps (10×10 μm$^2$) of the PbS nanocrystal film emission were recorded with Renishaw inVia micro-Raman microscope equipped with a 50× (0.75 N.A.) objective. Excitation wavelength was $\lambda_{ex}$= 785 nm. Step size in x and y direction was 0.4 μm.

ASSOCIATED CONTENT

**Supporting Information**
Plasmonic resonances of the rhombus
- Fig. S1. Eigenmodes of the rhombus, and resonant wavelength *versus* size.

Mathematical approach to obtain the aperiodic patterns with desired functionality
- Fig. S2. SEM images and calculated diffraction patterns of the metasurfaces.

Polarized microscopy experiments on the TBE pattern
- Fig. S3. Polarized microscopy images of the TBE structures.

Experimental details of the TPPL experiments
- Fig. S4. TPPL intensity *versus* incident laser power.
- Fig. S5. Scheme of the TPPL setup.

Normalized and vertical components of the electric field
- Fig. S6. COMSOL modeling of the electric field of the TBE structure.
- Fig. S7. Calculated extinction, scattering and reflection cross section, and integrated experimental TPPL intensity of the TBE pattern.

<u>Wavelength and size dependence of the optical near-field modulation of the TBE structure, and size scaling of its elements.</u>
- Fig. S8. Size, wavelength and polarization dependence of the TPPL signal of the TBE pattern.
- Fig. S9 – Element size (long axis D) *versus* pattern diameter for the TBE structures.

<u>Properties of the PbS nanocrystals.</u>
- Fig. S10. Transmission electron microscopy image and absorbance and emission spectra of the PbS nanocrystals.

AUTHOR INFORMATION

**Corresponding Author**

*roman.krahne@iit.it

**Author Contributions**

The scientific concepts and experimental designs were developed by M.M, R.K., N.J.B., A.W.B., and P.J.S. M.M. fabricated the structures tutored by D.S., performed the finite elements simulations with the assistance of R.P., and conducted the TPPL experiments under guidance of N.J.B. B.M.G. performed the synthesis of PbS nanocrystals. B.M.G. and D.S. fabricated and characterized the PbS nanocrystal films. M.M. and R.K. wrote the paper with contributions from all authors.

ACKNOWLEDGMENT

The research leading to these results has received funding from Horizon 2020 under the Marie Skłodowska-Curie Grant Agreement COMPASS No. 691185, and from the DOE through the Molecular User project #4754. Work at the Molecular Foundry was supported by the Director, Office of Science, Office of Basic Energy Sciences, Division of Materials Sciences and Engineering, of the U.S. Department of Energy under Contract No. DE-AC02-05CH11231.

B.M.G. acknowledges funding from the European Union's Horizon 2020 research and innovation program under grant agreement no.785219-GrapheneCore2. We thank the Clean Room Facility of the Italian Institute of Technology for support in device fabrication.

**TOC image**

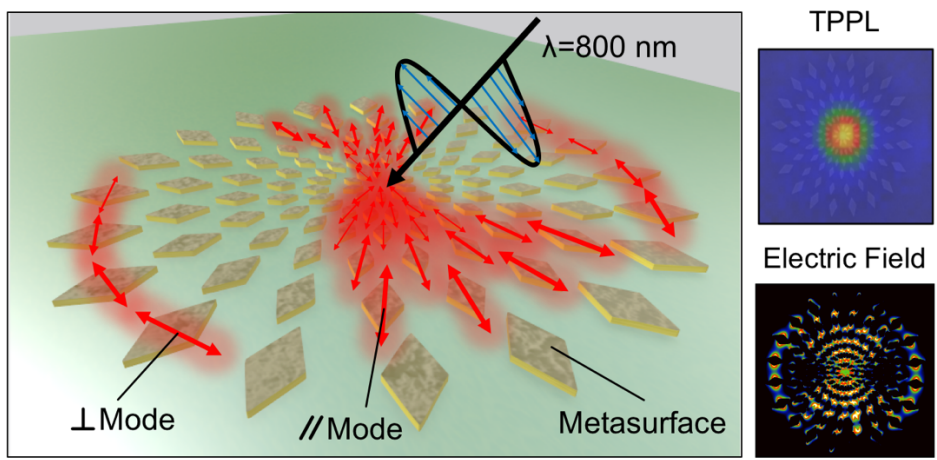